\documentclass[twocolumn,english,reprint, longbibliography, superscriptaddress, breaklinks=true, showkeys, showpacs=false, nofootinbib]{revtex4}
\usepackage[T1]{fontenc}
\usepackage[latin1]{inputenc}
\usepackage{color}
\usepackage{babel}
\usepackage{amsmath}
\usepackage{amssymb}
\usepackage{subfigure}
\usepackage{graphicx}
\usepackage{physics}
\usepackage{subfigure}

\newtheorem{definition}{Definition}
\newtheorem{lemma}{Lemma}
\newcommand{\argmin}{\arg\!\min}
\usepackage[unicode=true,pdfusetitle,
 bookmarks=true,bookmarksnumbered=false,bookmarksopen=false,breaklinks=true,pdfborder={0 0 0},backref=false,colorlinks=true]
 {hyperref} 
\makeatletter
\@ifundefined{textcolor}{}
{%
 \definecolor{BLACK}{gray}{0}
 \definecolor{WHITE}{gray}{1}
 \definecolor{RED}{rgb}{1,0,0}
 \definecolor{GREEN}{rgb}{0,1,0}
 \definecolor{BLUE}{rgb}{0,0,1}
 \definecolor{CYAN}{cmyk}{1,0,0,0}
 \definecolor{MAGENTA}{cmyk}{0,1,0,0}
 \definecolor{YELLOW}{cmyk}{0,0,1,0}
}
\pdfoutput=1
\hypersetup{colorlinks=true,citecolor=blue,linkcolor=cyan,urlcolor=blue,filecolor= green, breaklinks=true}
\usepackage{url}
\usepackage{breakurl}
\makeatother

\begin{document}

\title{Avoiding barren plateaus with classical deep neural networks}

\author{Lucas Friedrich}
\email[Electronic address: ]{lucas.friedrich@acad.ufsm.br}
\affiliation{Physics Departament, Center for Natural and Exact Sciences, Federal University of Santa Maria, Roraima Avenue 1000, 97105-900, Santa Maria, RS, Brazil}

\author{Jonas Maziero}
\email[Electronic address: ]{jonas.maziero@ufsm.br}
\affiliation{Physics Departament, Center for Natural and Exact Sciences, Federal University of Santa Maria, Roraima Avenue 1000, 97105-900, Santa Maria, RS, Brazil}

\selectlanguage{english}%

\begin{abstract}

Variational quantum algorithms (VQAs) are among the most promising algorithms in the era of Noisy Intermediate Scale Quantum Devices. Such algorithms are constructed using a parameterization U($\pmb{\theta}$) with a classical optimizer that updates the parameters $\pmb{\theta}$ in order to minimize a cost function $C$. For this task, in general the gradient descent method, or one of its variants, is used. This is a method where the circuit parameters are updated iteratively using the cost function gradient. However, several works in the literature have shown that this method suffers from a phenomenon known as the Barren Plateaus (BP). In this work, we propose a new method to mitigate BPs. In general, the parameters $\pmb{\theta}$ used in the parameterization $U$ are randomly generated. In our method they are obtained from a classical neural network (CNN). We show that this method, besides to being able to mitigate BPs during startup, is also able to mitigate the effect of BPs during the VQA training. In addition, we also show how this method behaves for different CNN architectures.

\end{abstract}

\keywords{ Variational Quantum Algorithms; Barren Plateaus; Neural Networks}

\maketitle

\section{Introduction}

With the rapid development of quantum computers (QCs), several applications are being proposed for them. Such devices use qubits as a building block, which are the quantum analogues of classical computing bits. Due to their properties, such as superposition and entanglement, QCs potentially have greater computing power than classical supercomputers. It is expected that several areas of knowledge will benefit from these devices, such as for example the simulation of quantum systems \cite{quantum_simulation}, the solution of linear systems of equations \cite{linear_system}, natural language processing \cite{quantum_nlp_1,quantum_nlp_2}, and the discovery of new drugs \cite{drug_discovery}.

Variational quantum algorithms (VQAs) \cite{VQA} are among the most promising computing strategies in the era of noisy intermediate scale quantum devices (NISQ). This era is characterized by the limited number of qubits we have access to and the presence of noise. These algorithms are built using a quantum part and a classical part. The quantum part in general is constructed using a parameterization V, which aims to take some data $\pmb{x}$ into a quantum state, followed by a parameterization $U(\pmb{\theta})$ with parameters $\pmb{\theta}$. Finally, measurements are performed that can be used to compute a cost function $C$ or as the input to another VQA or to a classical layer. This second case is widely used in quantum-classical hybrid neural networks (HQCNN). Finally, the classical part of the VQAs is responsible for optimizing the parameters $\pmb{\theta}$ in order to minimize or maximize $C$. In general, to perform this optimization we use the gradient descent method.

Despite the benefits of VQAs, some issues still need to be resolved before we can use them to their full capacity. One of the most important problems that this method faces is the phenomenon known as gradient vanishing, or Barren Plateaus (BP). This phenomenon is characterized by the flattening of the cost function landscape. Because of this, the number of times that the cost function $C$ must be evaluated in order to train the parameters $\pmb{\theta}$ grows exponentially with the number of qubits.

Some results from the literature show that this phenomenon is also related to the choice of cost function \cite{BR_cost_Dependent}, entanglement \cite{BR_Entanglement_devised_barren_plateau_mitigation,BR_Entanglement_induced_barren_plateaus}, parameterization expressivity \cite{BR_expressibility}, and the presence of noise \cite{BR_noise} . Furthermore, it has also been shown that gradient-free methods may suffer from BP  \cite{BR_gradientFree}. Because of this, some methods have been proposed to alleviate BP, such as parameter initialization strategies \cite{BR_initialization_strategy}, parameter correlation \cite{BR_Large_gradients_via_correlation}, pre-training \cite{BR_LSTM}, and layer-by-layer training \cite{BR_layer_by_layer}.

In this article, we report an alternative method for avoiding BP. Our method consists of using a classical neural network for producing the initial parameters $\pmb{\theta}$. We show that this is a simple and viable method that can be used in order to escape from the BP problem.

The remainder of this article is organized as follows. In Sec. \ref{SC_VQA}, we review quantum variational circuits, discussing data encoding in quantum states in Sec. \ref{data_encoder}. The choice of parameterization is dealt with in Sec. \ref{Parametrization} and the cost function is described in Sec. \ref{cost_function}. Afterwards, in Sec. \ref{training}, we discuss the training process for VQAs. Next, in Sec. \ref{Barren_Plateaus}, we address the problem of Barren Plateus and its consequences.  After that, we present the proposed method in Sec. \ref{method}. In Sec. \ref{result} we report our numerical results. Finally, we present our conclusions in Sec. \ref{conclusion}.

\begin{figure}
    \centering
    \includegraphics[scale=0.5]{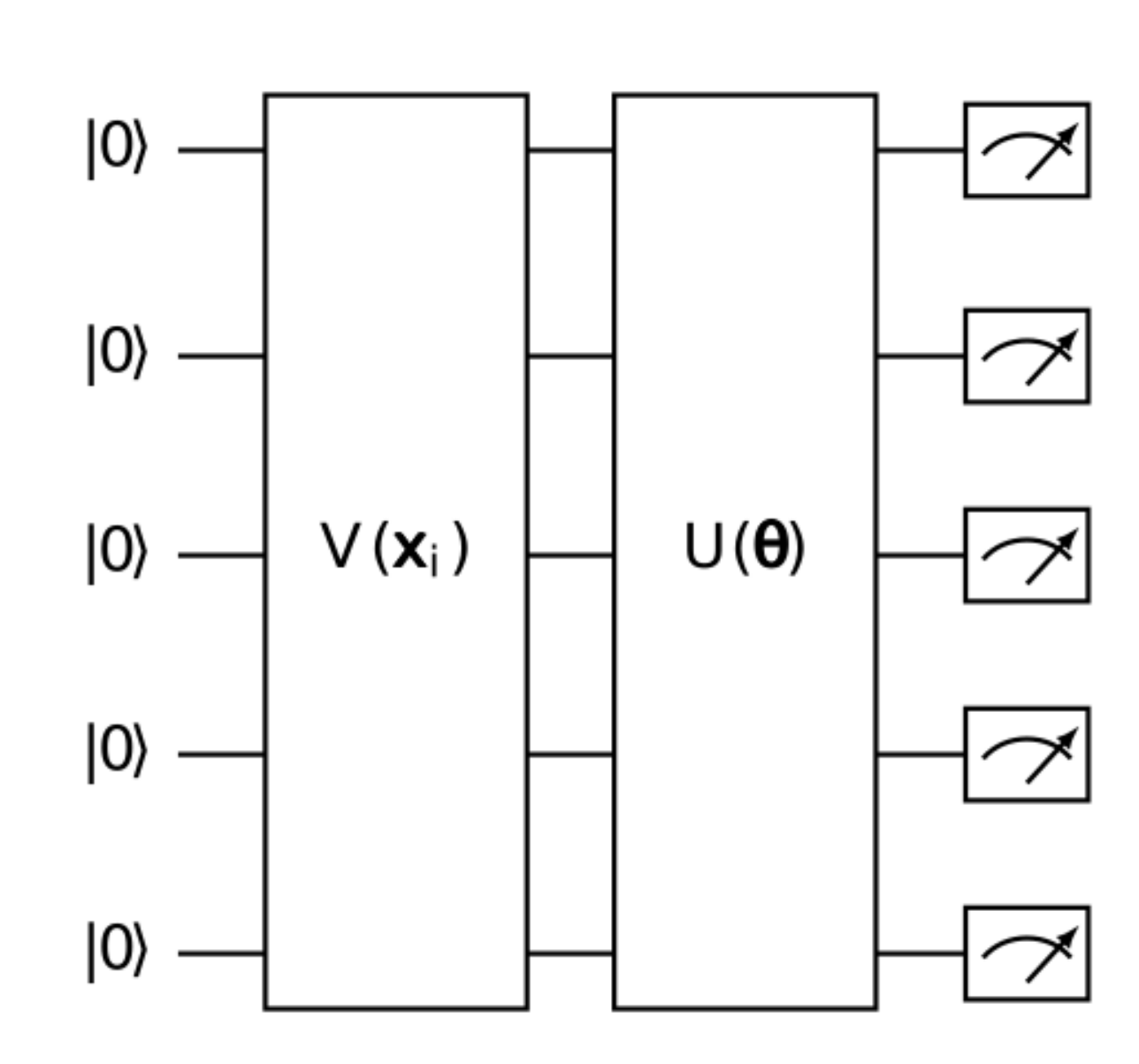}
    \caption{In this figure is illustrated the quantum part of Variational Quantum Algorithms (VQAs). For this circuit, we used five qubits, all starting in the $|0\rangle$ state. We can see, from the figure, that VQAs have two sets of unitary transformations. The first unitary, $V$, is used to encode the data in a quantum state. The second unitary, $U(\pmb{\theta})$, is an arbitrary parameterization with parameters $\pmb{\theta}$. These parameters are updated iteractively in order to minimize a cost function $C$.}
    \label{fig:VQA_illustracao}
\end{figure}

\section{Variational quantum algorithms}
\label{SC_VQA}

In this section we give a quick review about VQAs. For this, we consider the following problem. Given a dataset $\mathcal{D} := \{ \pmb{x}_{i}, \pmb{y}_{i} \}_{i=1}^{N} $, we want to create a quantum circuit that, given the input $x_{i}$ returns its respective output $y_{i}$. For this, we must follow three steps. First we must map our data $x_{i}$ into a quantum state. After that, we must apply a parameterization U($\pmb{\theta}$). Finally, we must perform measurements to calculate the cost function $C$. In the next three subsections we will briefly describe each of these steps.

\subsection{Encoder}
\label{data_encoder}

Given the  dataset $\mathcal{D}$, the first task is to map the data into a quantum state. That is, we must prepare a state $|\pmb{x}_{i} \rangle = V(\pmb{x}_{i})| 0 \rangle $, with $| 0 \rangle = | 0 \rangle^{\otimes n}$, where $n$ is the number of qubits and $V(\pmb{x}_{i})$ is the parameterization that will take the data into a quantum state. The parameterization can take different forms \cite{data_encoder_1,data_encoder_2,data_encoder_3}, below are some definitions for V.

\begin{definition}
(Wavefunction Encoding). Given a vector $\pmb{x}_{i} \in \mathbb{R}^{N} $, the data is encoded in the wave function, that is to say, we prepare the state
\begin{equation}
    | \pmb{x}_{i} \rangle  = \frac{1}{ \| \pmb{x}_{i} \|_{2}^{2} }\sum_{j=1}^{N} x_{i}^{j}|j \rangle
\end{equation}
where $N=2^{n}$. So, the initial density matrix is  $\rho_{init} = V(\pmb{x}_{i})| 0 \rangle \langle 0|V(\pmb{x}_{i})^{\dagger}  = | \pmb{x}_{i} \rangle \langle \pmb{x}_{i} |$.
\label{df:1}
\end{definition}

\begin{definition}
(Qubit Encoding). In Qubit Encoding, each qubit encodes one data value, that is, given a vector $\pmb{x}_{i} \in \mathbb{R}^{n} $, we prepare the state
\begin{equation}
    | \pmb{x}_{i} \rangle  = \bigotimes_{j=1}^{n} \cos(  x_{i}^{j} )|0 \rangle + \sin(  x_{i}^{j} )|1 \rangle.
\end{equation}
So, the initial density matrix is again $\rho_{init} = V(\pmb{x}_{i})| 0 \rangle \langle 0|V(\pmb{x}_{i})^{\dagger}  = | \pmb{x}_{i} \rangle \langle \pmb{x}_{i} |$.
\label{df:2}
\end{definition}

These are some of the most used forms of data encoding in the literature, and the choice will depend on the problem at hand.

\subsection{Parameterization}
\label{Parametrization}

The parameterization and the encoding have great influence on the performance of VQAs. The choice of parameterization depends on the task to which it will be used. In general, this is given by a parameterization $U$ with parameters $\pmb{\theta}$. In VQAs, these parameters are updated iteractively in order to minimize or maximize a cost function $C$. The choice between minimizing or maximizing $C$ will depend on the task to which the VQAs will be applied. 

A possible parameterization is
\begin{equation}
    U(\pmb{\theta}) = \prod_{i=1}^{L}U(\pmb{\theta}_{i})W_{i} \label{eq:paramtrization_1}
\end{equation}
with
\begin{equation}
    U(\pmb{\theta}_{i}) = \bigotimes_{j=1}^{n} R_{\sigma}(\theta_{i}^{j}), \label{eq:paramtrization_2}
\end{equation}
where the indices $i$ and $j$ represent the layer and the qubit, respectively. Besides, $ R_{\sigma}(\theta_{i}^{j}) = e^{-i \theta_{i}^{j} \sigma/2} $ with $\sigma \in ( \sigma_{x},\sigma_{y},\sigma_{z}, ) $ being one of the Pauli matrices.
The unitaries $W_{i}$ are unparameterized units, that is, they do not depend on $\pmb{\theta}$.
The Eq. (\ref{eq:paramtrization_1}) is a parameterization widely used in Quantum Neural Networks (QNNs), being considered an analogue to the classical models of Deep Neural Networks (DNNs).

As there are various ways of constructing the parameterization $U(\pmb{\theta})$, an immediate question is about what is the best construction given a certain task, as for example to prepare an arbitrary final state $|y\rangle$. For this, we can consider the expressivity of the parameterization \cite{expressibility_1,expressibility_2}. Expressivity is defined as the ability of the parameterization to access the Hilbert space. Therefore, the more expressive a parameterization is, the more it can access the states of the Hilbert space. Hence the greater is the probability of getting the state $|y\rangle$. However, as shown in Ref. \cite{Expressibility_barren_plateaus}, there is a relationship between expressivity and BP, where the higher the expressivity the more the VQA suffers from the BP phenomenon.

\subsection{Cost Function}
\label{cost_function}
A fundamental aspect of VQAs is the cost function, as it is used to optimize the model parameters. The choice of cost function depends on the task. For example, if we are working with problems where we are given a dataset $\mathcal{D}$, where our goal is given $\pmb{x}_{i}$ to get the corresponding output $\pmb{y}_{i }$, a possible choice for the cost function is
\begin{equation}
    C = \frac{1}{N}\sum_{i=1}^{N}Tr[\rho_{i}H_{i}]
\end{equation}
where
\begin{equation}
    \rho_{i} = U(\pmb{\theta})|\pmb{x}_{i} \rangle  \langle \pmb{x}_{i}| U(\pmb{\theta})^{\dagger}
\end{equation}
and
\begin{equation}
    H_{i} = |\pmb{y}_{i} \rangle  \langle \pmb{y}_{i}|,
\end{equation}
with $|\pmb{y}_{i} \rangle$ being the quantum state encoding $\pmb{y}_{i}$.

We can also consider the case of quantum-classical hybrid neural networks (HQCNN), where we create such models using quantum and classical layers together. In this case, we can have the output of the quantum layer being used as input for the classical layer. Then we can define the cost function as
\begin{equation}
    C = \frac{1}{N}\sum_{i=1}^{N}L( \overline{\pmb{y}}_{i}  ,\pmb{y}_{i} ),
\end{equation}
where L is a function that compares the desired output $\pmb{y}_{i}$ with the output $\overline{\pmb{y}}_{i}$ obtained from the HQCNN.

\section{Training}
\label{training}

Given the data set $\mathcal{D}$ and defined the encoding unitary $V$, parameterization $U$, Eq. \eqref{eq:paramtrization_1} with parameters $\pmb{\theta}$, and the cost function $C$, we aim to find 
\begin{equation}
    \pmb{\theta}^{*} = \argmin_{\pmb{\theta}} C(\pmb{\theta}),
\end{equation}
where $\pmb{\theta}^{*}$ are the optimal parameters. For this, we must train/adjust the quantum circuit. To perform the training, in general the gradient descent method is used. This method consists of an iteration where the gradient of the cost function $C$ is used to update the parameters, that is
\begin{equation}
    \pmb{\theta}^{t+1} = \pmb{\theta}^{t}-\eta \nabla_{\pmb{\theta}}C,
    \label{eq:gradient}
\end{equation}
where $\eta$ is the learning rate and $t$ is the epoch.

From Eq. \eqref{eq:gradient}, we see that the training depends on the value of the derivatives of the cost function with respect to the parameters $\pmb{\theta}$. Here we have to mention two problems that this method has. The first problem happens when the values of the derivatives are too large. In this case, the issue that can occur is that when we update the parameters, we pass through the global minimum, which is the point that we want to reach, and get stuck in a local minimum that is not ideal for $\pmb{\theta}^{* }$. To get around this problem, we can simply use a small value for $\eta$. The second problem happens if the derivative values are too small. In this case the updating of the parameters will be extremely slow, that is, the number of epochs needed to train the parameters $\pmb{\theta}$ will increase very rapidly with the number of parameters in the quantum circuit. Next, we will see that the phenomenon known as Barren Plateaus (BP) is related to this second problem.

\section{Barren Plateaus}
\label{Barren_Plateaus}
For the purposes of this section, we will rewrite Eq. \eqref{eq:paramtrization_1} as follows
\begin{equation}
    U(\pmb{\theta}) = U_{L}(\pmb{\theta})U_{R}(\pmb{\theta}) \label{eq:parametrizarion_br}
\end{equation}
with
\begin{equation}
    U_{R}(\pmb{\theta}) = \prod_{i=1}^{k}U(\pmb{\theta}_{i})W_{i}  \label{eq:parametrizarion_br_r}
\end{equation}
and
\begin{equation}
    U_{L}(\pmb{\theta}) = \prod_{i=k+1}^{L}U(\pmb{\theta}_{i})W_{i} \label{eq:parametrizarion_br_l}
\end{equation}
Furthermore, we will define the cost function as
\begin{equation}
    C = Tr[HU(\pmb{\theta}) \rho U(\pmb{\theta})^{\dagger}] \label{eq:cost_function_geral}
\end{equation}
where $\rho$ is the initial state and $H$ is a Hermitian operator.

Let the cost function $C$ be defined in Eq. \eqref{eq:cost_function_geral} with parameterization $U$ given as in Eq. \eqref{eq:parametrizarion_br}. If $U_{R}$, Eq. \eqref{eq:parametrizarion_br_r}, and $U_{L}$, Eq. \eqref{eq:parametrizarion_br_l}, are $2$-designs, then the average value of the partial derivative of $C$ with respect to a parameter $\theta_{k}$ will be
\begin{equation}
  \langle  \partial_{k}C \rangle_{U} = 0. \label{eq:var_n}
\end{equation}
Here we use the notation $ \frac{\partial C}{\partial \theta_{k}} := \partial_{k}C $. With this, we see that the gradients of the cost function $C$ are not biased in any direction. Also, since $\langle \partial_{k}C \rangle_{U} = 0$, we have that
\begin{equation}
    Var(\partial_{k}C) = \langle(\partial_{k}C)^{2} \rangle.
\end{equation}
The proof of Eq. \eqref{eq:var_n} is presented in the Appendix.

\begin{definition}\label{tr:1}
(Barren Plateaus). Let the cost function $C$ be defined as in Eq. \eqref{eq:cost_function_geral} with parameterization $U$ given as in Eq. \eqref{eq:parametrizarion_br} with $U_{R}$, Eq. \eqref{eq:parametrizarion_br_r}, and $U_{L}$, Eq. \eqref{eq:parametrizarion_br_l}. The cost function exhibits a barren plateau if, for all $\theta_{k} \in \pmb{\theta}$, the variance vanishes exponentially with the number of qubits $n$ as

\begin{equation}
    Var[\partial_{k}C] \leqslant   F(n), \text{ with } F(n) \in \mathcal{O}\bigg( \frac{1}{b^{n}} \bigg) \label{eq:variance}
\end{equation}
for all $b>1$.
\end{definition}

Based on Chebyshev's inequality,
\begin{equation}
    Pr( |\partial_{k}C| \geqslant \delta )  \leqslant \frac{ Var(\langle \partial_{k}C \rangle)  }{\delta^{2}}, \label{eq:Chebyshev}
\end{equation}
we see that the probability that $\partial_{k}C$ deviates from its mean, $ \langle \partial_{k}C \rangle_{U} = 0$, is less than or equal to the variance, which in this case is given by Eq. \eqref{eq:variance}. So, we see from Definition 3 that this probability decreases as the number of qubits increases. Therefore, for quantum circuits that use several qubits, on average the value of $\partial_{k}C$ will be equal to or close to zero. With this, we see that quantum circuits naturally suffer from the second problem concerning the gradient method. Therefore, in general the number of times that the cost function, Eq. \eqref{eq:cost_function_geral}, must be evaluated in order to obtain $\pmb{\theta}^{*}$ shall grow exponentially with the number of qubits.

This phenomenon is known as Barren Plateaus (BP). It was initially observed in Ref. \cite{Barren_Plateaus_1}, where its dependence on the number of qubits and parameterization depth was shown. In Ref. \cite{BR_cost_Dependent}, the authors extended the results obtained in Ref. \cite{Barren_Plateaus_1}, showing that the BP are also related to the choice of the cost function. In addition, BP are also related to entanglement \cite{BR_Entanglement_devised_barren_plateau_mitigation,BR_Entanglement_induced_barren_plateaus}, expressivity \cite{BR_expressibility}, presence of noise \cite{BR_noise}, and gradient free methods \cite{BR_gradientFree}.

Some methods have been proposed with the aim of overcoming the BP, such as parameter initialization strategies \cite{BR_initialization_strategy}, correlated parameters \cite{BR_Large_gradients_via_correlation}, pre-training \cite{BR_LSTM}, and layer-by-layer training \cite{BR_layer_by_layer}. In the next section, we show how Classical Deep Neural Networks can be used for dealing with the BP problem.

\section{Method}
\label{method}

\begin{figure}
    \centering
    \includegraphics[width=90mm]{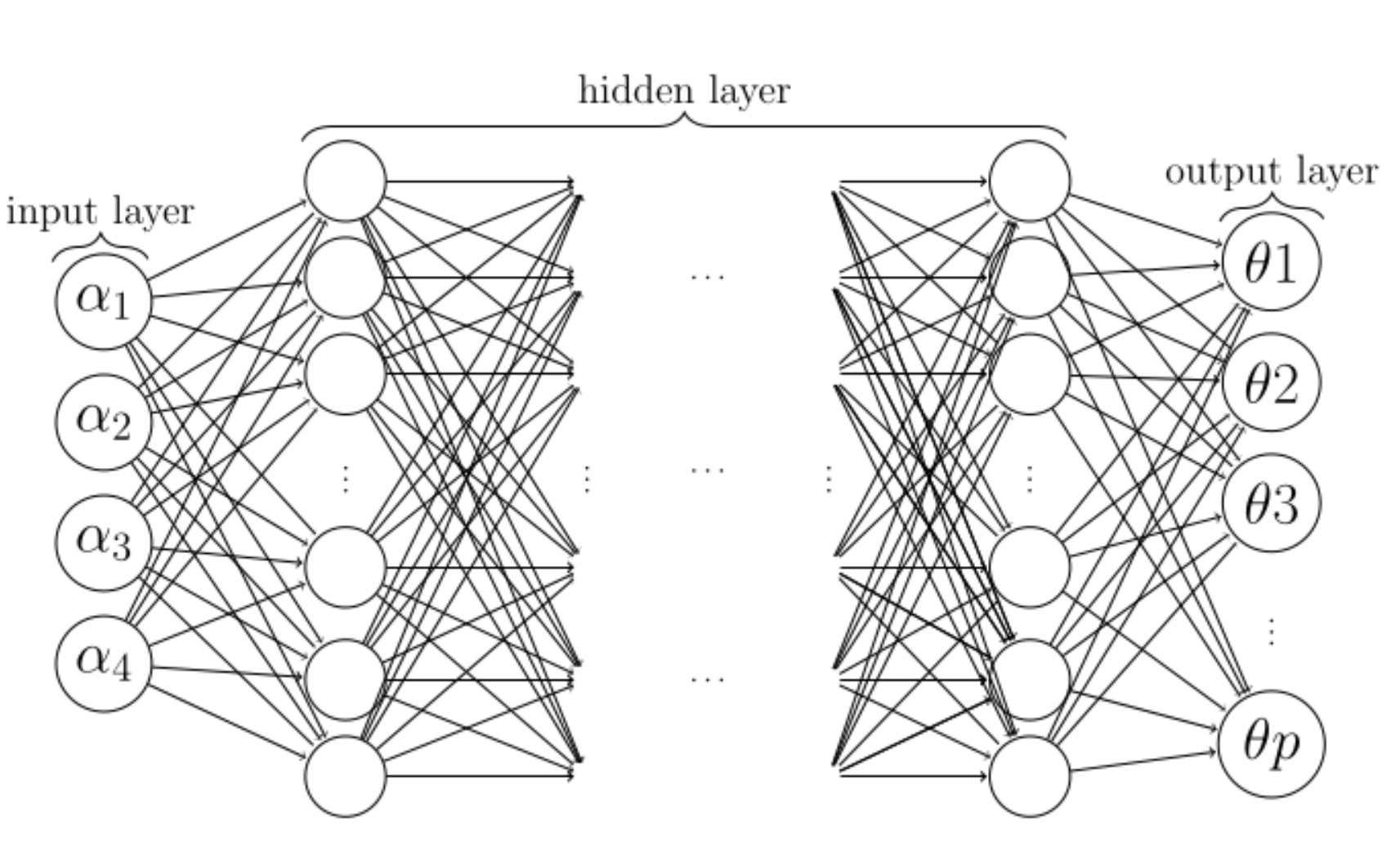}
    \caption{Model of a deep neural network. In the input layer, the vector of variables $\pmb{\alpha}$ is passed. This vector is optimized along with the network parameters. In this figure, the number of hidden layers is not specified. In the output layer, a vector of dimension $p$ is returned. This dimension is equal to the number of parameters of the quantum circuit. In this case $p=Ln$, where $n$ is the number of qubits used and $L$ is the parameterization depth, given as in Eq. \eqref{eq:paramtrization_1} and illustrated in Fig. \ref{fig:parametrization1000}. The values of this output vector will be used as the parameters for the parameterization, given in Eq. \eqref{eq:paramtrization_1}, of the quantum circuit.}
    \label{fig:deep_neural_network}
\end{figure}

In this section we will describe the proposed method. Initially we must remember that given a VQA, Fig. \ref{fig:VQA_illustracao}, in general the parameters $\pmb{\theta}$ are randomly generated from a probability distribution and are applied directly to the quantum circuit. During training, they are updated using the gradient of the cost function. As we discussed above, results from the literature show that this initialization method results in BP. In this work, we propose that the parameters $\pmb{\theta}$ that will be used in the quantum circuit are not to be randomly generated and used directly as initial parameters for the quantum circuit. Instead, we generate a vector of $m$ elements, which we denote by $\pmb{\alpha}$. This vector is used as input to a Deep Neural Network (DNN), Fig. \ref{fig:deep_neural_network}. In these numerical simulation, we used $m=4$. The output of this network, which will be a vector with dimension equal to the number of parameters of the quantum circuit, will be used as the input parameters $\pmb{\theta}$. Then, the cost function will have the form
\begin{equation}
    C = Tr[HU(\pmb{\theta}(\phi)) \rho U(\pmb{\theta}(\phi))^{\dagger}],
\end{equation}
with $\phi$ being the DNN parameters.


During training the $\phi$ parameter will be updated using gradient descent. Furthermore, it should be noted that the parameters of the vector $\pmb{\alpha}$ can either be updated using gradient descent or can be considered as constant values. For this work we chose the first case.

Furthermore, training a parameterized quantum circuit (PQC) on a quantum computer using this scheme is similar to training a classical-quantum hybrid neural network. But instead of using the output of the classical layer as input to the quantum layer, it will be used as a parameter of the quantum layer.  Or, to put it in another way, we can consider the output of the quantum layer as a function $F$ that depends on the classical layer, which is also a function that depends on the parameters $\phi$ , let say G($\phi$). So, we have that the output will be $y = F(G(\phi))$. Thus the training of a PQC consists of optimizing the parameters $\phi$. For this we use the derivatives with respect to these parameters. Once $y = F(G (\phi))$, we must use the chain rule to get the derivatives. However, we should note that $F$ is the quantum part. Therefore, when we use the chain rule, the derivatives of the quantum part will be obtained by the parameter-shift-rule method. 

For our numerical simulation, we consider a VQA similar to the one shown in Fig. \ref{fig:VQA_illustracao}, with input data $\pmb{x} = (\frac{\pi}{4},\frac{\pi}{4}, . . . , \frac{\pi} {4}) $ that will be encoded into a quantum state using the definition \ref{df:2}. The parameterization $U$ will be given by Eq. \eqref{eq:paramtrization_1} with $\sigma=Y$ for all $R_{\sigma}$ that appear in Eq. \eqref{eq:paramtrization_2}.

\begin{figure}
    \centering
    \includegraphics[scale=0.4]{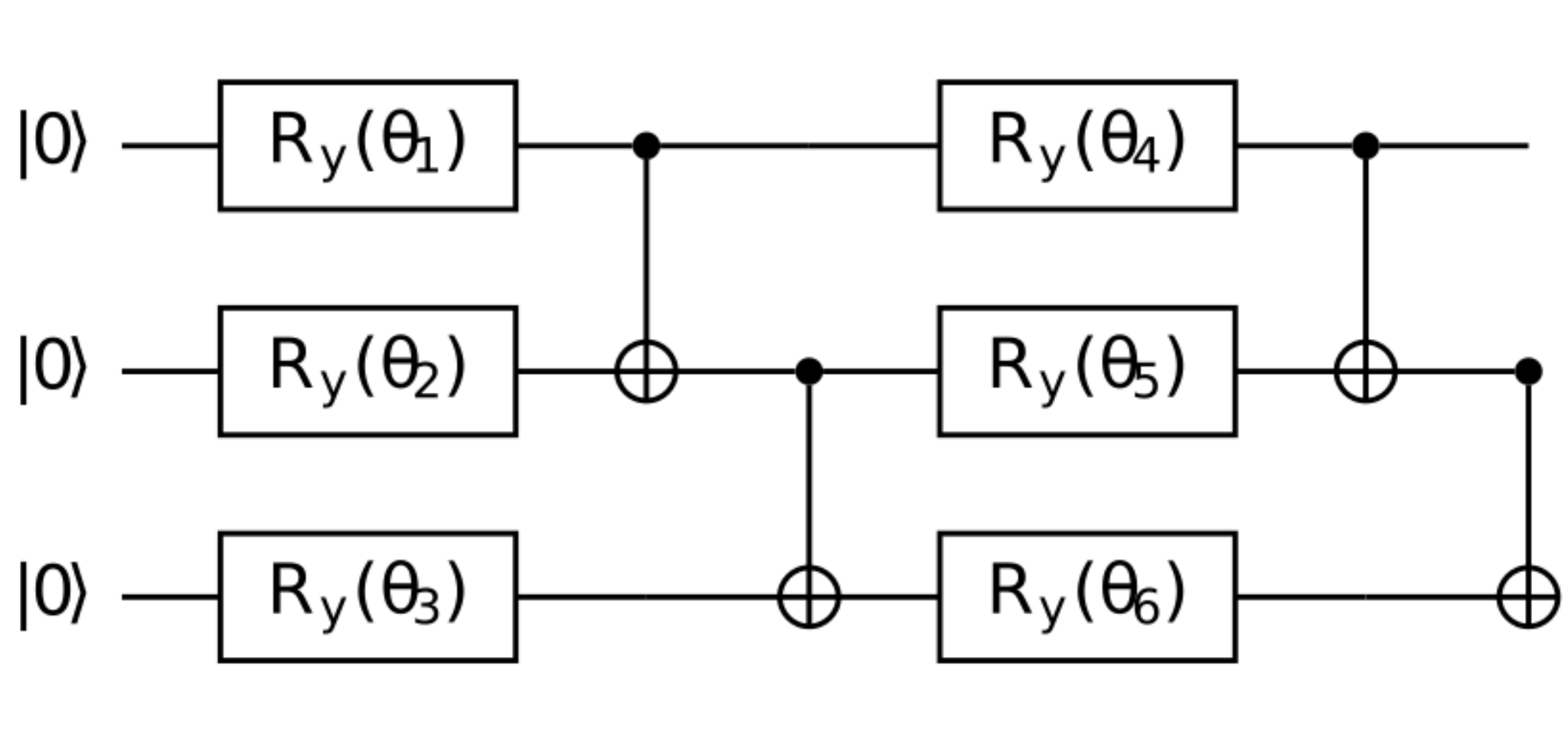}
    \caption{Example of how the parameterization defined in Eq. \eqref{eq:paramtrization_1} looks like for three qubits and two layers, i.e. , $L=2$. We choosed to use $\sigma=Y$ for all $R_{\sigma}$ that appear in Eq. \eqref{eq:paramtrization_1}. }
    \label{fig:parametrization1000}
\end{figure}

For all simulation, we define the cost function as
\begin{equation}
    C = 1- \frac{1}{n}\sum_{i}^{n} Tr\big[(| 0 \rangle \langle 0 |_{i} \otimes \mathbb{I}_{\overline{i}})\rho(\pmb{\theta}(\phi)) \big], \label{eq:cost_experiment}
\end{equation}
where the index $i$ in Eq. \eqref{eq:cost_experiment} indicates that $| 0 \rangle \langle 0 |$ acts on the qubit of the same index, $\overline{i}$ indicates that $\mathbb{I}$ acts on all qubits except the one with index $i$ and $\rho( \pmb{\theta}(\phi))$ is the final state. We showed in Definition \ref{tr:1} the definition of BP for cost functions in the form given by Eq. \eqref{eq:cost_function_geral}. The cost function defined in Eq. \eqref{eq:cost_experiment} is a generalization of Eq. \eqref{eq:cost_function_geral} and also suffers from BP. For more details, see Ref.  \cite{BR_cost_Dependent}.

There are two ways in which one can analyze whether a VQA suffers from BP. One is by computing the variance of the partial derivative of the $C$ function with respect to some parameter $\theta_{k}$. The second method is by counting how many epochs it takes to get a given value of $C$. That is to say, setting a fixed value for $C$, for example $C=0.1$, we count how many iterations are needed to optimize the parameters $\pmb{\theta}$ until we get $\pmb{\theta}^{*}$ using the rule defined in Eq. \eqref{eq:gradient}. In this case, if the VQA suffers from BP, the number of iterations required will grow exponentially with the number of qubits, as seen in Theorem \ref{tr:1}.

During the simulations, for each model considered, ten different simulations will be performed. That is, once defined the model, the number of qubits, and initialized the random variables, the model will be optimized until we obtain $\pmb{\theta}^{*}$. After this, new random variables will be generated and again the model will be optimized until obtaining $\pmb{\theta}^{*}$. This process is repeated ten times. After that, we calculate the average of epochs needed in order to obtain $\pmb{ \theta}^{*}$.

\section{Results}
\label{result}
\begin{figure}
\centering
    \includegraphics[scale=0.39]{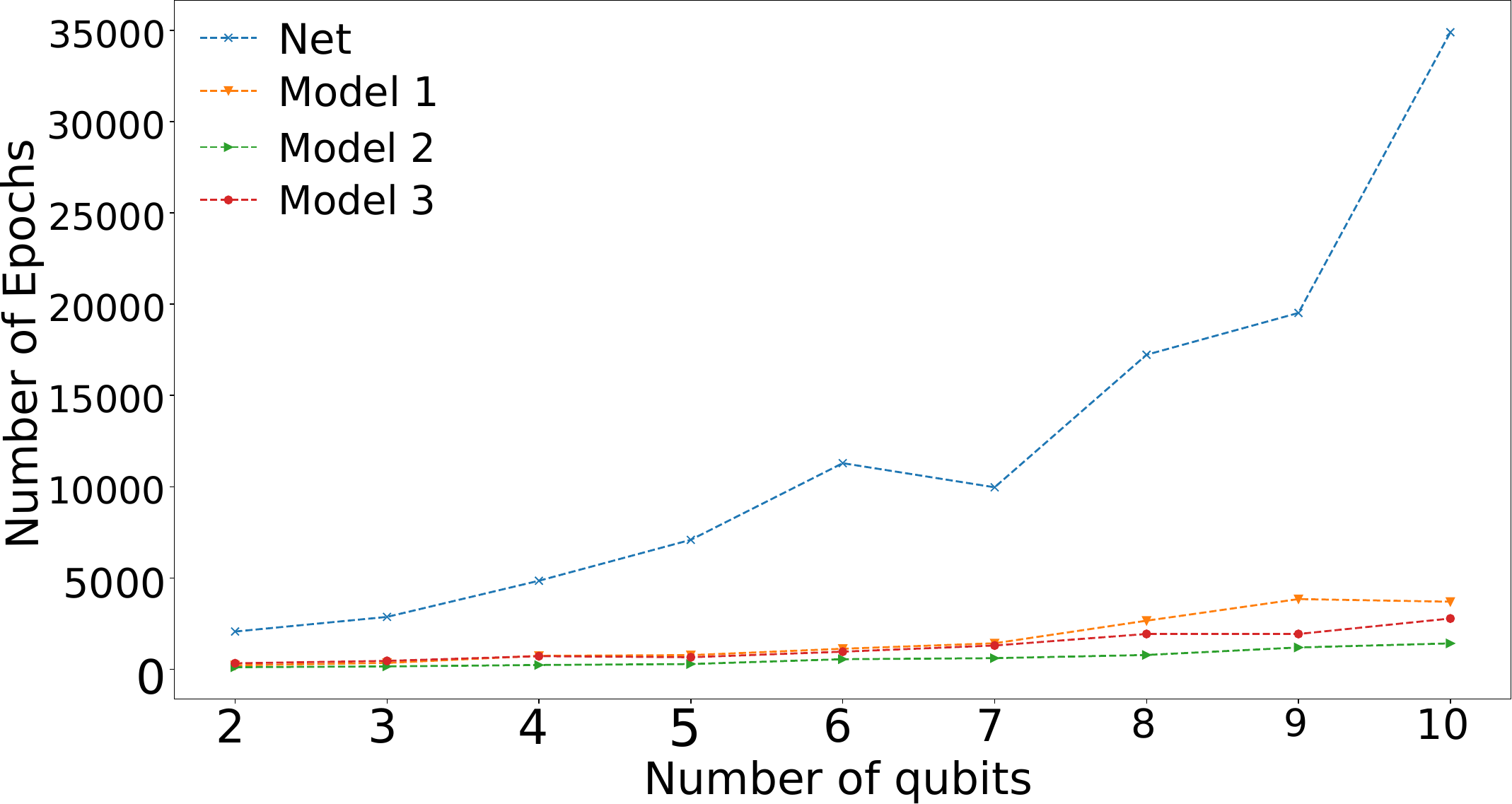}
    \caption{Number of epochs needed to obtain $\pmb{\theta}^{*}$, using Eq. \eqref{eq:cost_experiment}, as a function of the number of qubits used.}
    \label{fig:result_grafic}
\end{figure}

For the first simulations, Fig. \ref{fig:result_grafic}, the depth of the parameterization, Eq. \eqref{eq:paramtrization_1}, grows linearly with the number of qubits, that is, $L=n$. With that, we guarantee that our model will suffer from Barren Plateaus (BP) \cite{BR_cost_Dependent}. We see, as shown in Fig. \ref{fig:result_grafic} in the blue dashed line, that in fact this occurs. We also define that our goal is to reach $\pmb{\theta}^{*}$ such that $C=0.001$.

In our simulations, we used three different parameterizations for the Deep Neural Network (DNN), as shown in Table \ref{tb:table1}. With this we can analyze how different architectures of the DNN affect the emergence of BP.

\begin{table}
\centering
\begin{tabular}{|l|l|}
\hline
model 1 & \begin{tabular}[c]{@{}l@{}}Linear(4,10)\\         Tanh()\\ Linear(10,$n L$)\\         Tanh()\end{tabular}                                    \\ \hline
model 2 & \begin{tabular}[c]{@{}l@{}}Linear(4,30)\\        Tanh()\\ Linear(30,$n L$)\\         Tanh()\end{tabular}                                      \\ \hline
model 3  & \begin{tabular}[c]{@{}l@{}}Linear(4,10)\\         Tanh()\\ Linear(10,20)\\         Tanh()\\ Linear(20,$n L$)\\         Tanh()\end{tabular}   \\ \hline
\end{tabular}
\caption{The different architectures used for the Deep Neural Networks. Here $n$ is the number of qubits used and $L$ the depth of the parameterization $U$ defined in Eq. \eqref{eq:paramtrization_1}.}
\label{tb:table1}
\end{table}

In our simulations, different activation functions were used, however. It was observed however that only when using the hyperbolic tangent function the model was able to learn. With this, we see that when using this method to avoid  BP, we must always use the hyperbolic tangent function as activation function for the DNN.

In Fig. \ref{fig:result_grafic}, for obtaining the blue dashed line, labeled Net, the model VQA was initialized with parameters generated from a uniform distribution over $[0,2\pi]$. In this case, these parameters were used directly in the parameterization $U$, Eq. \eqref{eq:paramtrization_1}. We can see that as the number of qubits increased, the number of epochs needed to obtain $\pmb{\theta}^{*}$ also increases. With that we see that this method, where the parameters are randomly generated and used directly in $U$, suffers from BP.

To obtain the graphs referring to the DNN models 1, 2 and 3, especified in Table \ref{tb:table1}, the vector $\pmb{\alpha}$ was also generated from a uniform distribution over $[0,2\pi]$. For the creation of the DNN, we used Pytorch \cite{pytorch_ref}. So, the initialization of the variables was done using the standard initialization of Pytorch.


We can see in Fig. \ref{fig:result_grafic} that the number of epochs needed to optimize the parameters of models 1, 2 and 3 also grew with the increase in the number of qubits. However, this increase was much smaller than for the Net model which we know suffers from BP as shown in \cite{BR_cost_Dependent}, moreover, it should be noted that the number of variables in these models also had a higher growth than the Net model . Therefore, the number of epochs needed to optimize $\pmb{\theta}$ is expected to increase. From this analysis, we can conclude that using a DNN in the input data that we call $\pmb{\alpha}$, we were able to mitigate the BP problem.

For the next numerical simulations, we will again use the cost function defined in Eq. \eqref{eq:cost_experiment}. But in this case our goal is to reach $\pmb{\theta}^{*}$ such that $C=0.3$. We used a high value for C because our objective is to analyze the relationship between the number of epochs needed to obtain $\pmb{\theta}^{*}$ in relation to the size of the system, that is, the number of qubits and the depth of the parameterization $U$. We use again the models defined in table \ref{tb:table1}. For the next simulations, the parameterization depth will not grow linearly with the number of qubits as in the previous simulations. Instead, we define a fixed size for this parameterization.

\begin{figure}[h]
\centering
    \includegraphics[scale=0.2]{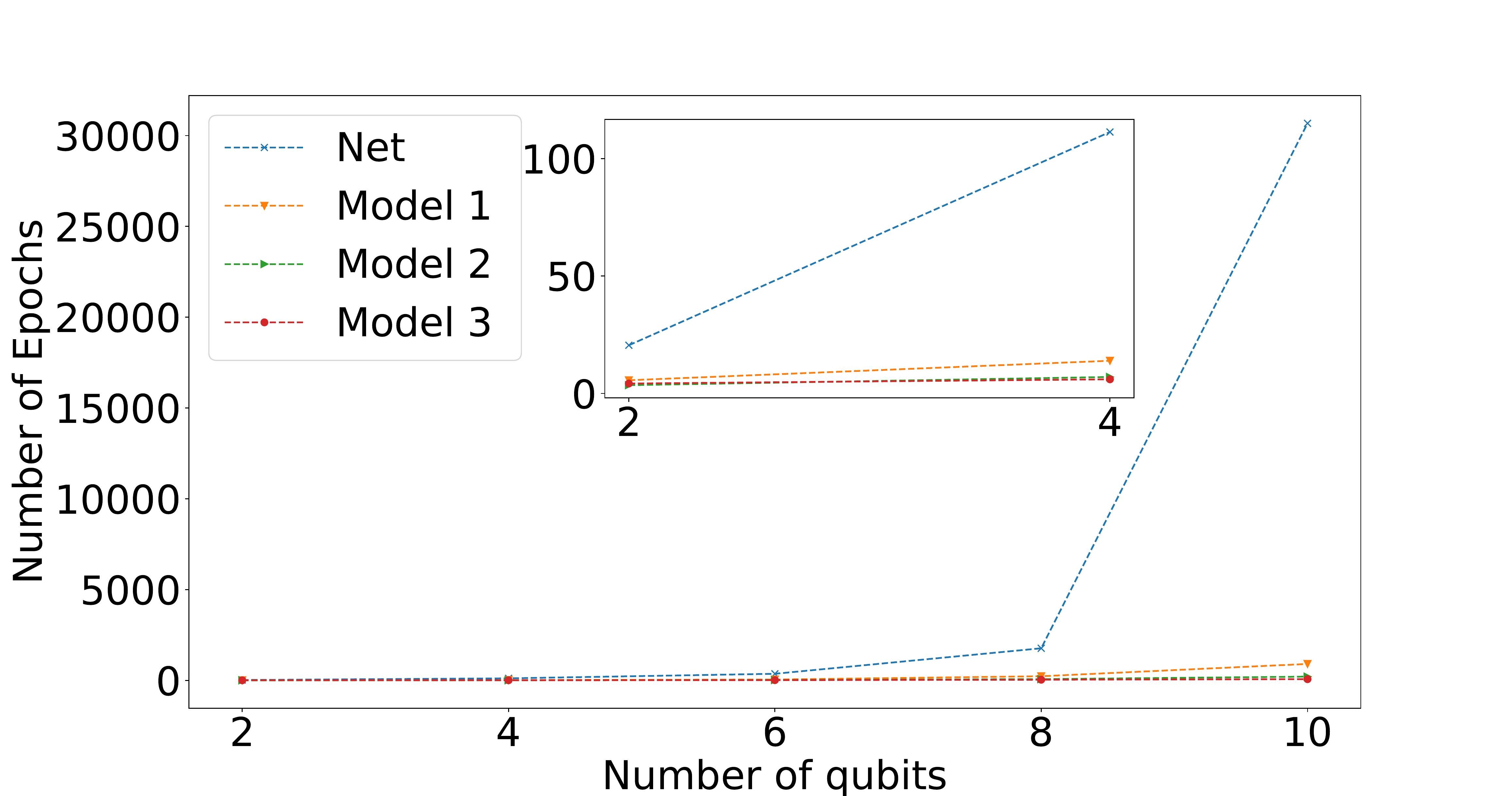}
    \caption{Number of epochs needed to obtain $\pmb{\theta}^{*}$, using Eq. \eqref{eq:cost_experiment}, as a function of the number of qubits used. In this case, we used $L=20$. }
    \label{fig:result_grafic_nl_20}
\end{figure}

\begin{figure}[h]
\centering
    \includegraphics[scale=0.2]{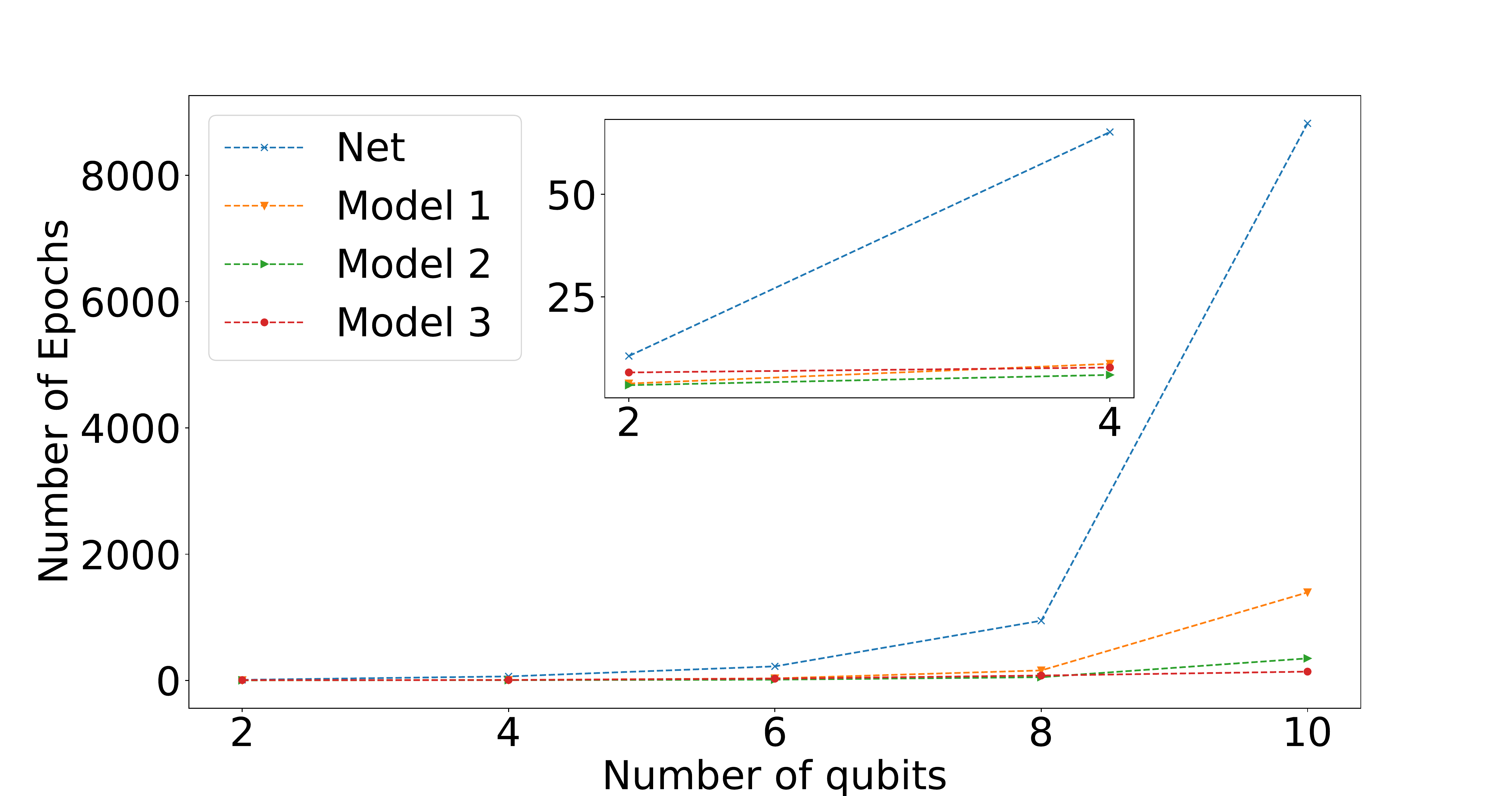}
    \caption{Number of epochs needed to obtain $\pmb{\theta}^{*}$, using Eq. \eqref{eq:cost_experiment}, as a function of the number of qubits used. In this case, we used $L=30$. }
    \label{fig:result_grafic_nl_30}
\end{figure}

We can see from Fig. \ref{fig:result_grafic_nl_20} that the Net model again suffers from BP. The models 1, 2 and 3 manage to avoid these problems. For the next numerical simulations, whose results are shown in Fig. \ref{fig:result_grafic_nl_30}, we use $L=30$ for the parameterization depth. Again, if we observe the behavior of the number of epochs needed to optimize the Net model, which suffers from BP, with the other three, we will see that they managed to alleviate the BP. It should be noted that for deep circuits the BP phenomenon does not depend on the cost function, as seen in \cite{BR_cost_Dependent}.

If we compare the results of Fig. \ref{fig:result_grafic_nl_20} and Fig. \ref{fig:result_grafic_nl_30}, we see that the number of epochs needed to obtain $\pmb{\theta}^{*}$ was smaller when using $L=30$. At first we would expect the number to be smaller when using $L=20$. With that, we conclude that for $L=30$ the initialization of the parameters $\pmb{\theta}$ was a little better than for $L=20$. However, we must always observe the behavior of the number of epochs needed to obtain $\pmb{\theta}^{*}$ with the size of the system. So we see that for $L=30$ the Net model also suffers from BP. Furthermore, we see that models 1, 2 and 3 again manage to avoid BP. However, for $n=10$ the performance of model 1 is inferior in relation to the other two models.

\section{Conclusion}
\label{conclusion}

In this work, we aimed to demonstrate that when we use a Deep Neural Network (DNN) in the input variables of a Variational Quantum Algorithm (VQA), we are able to avoid the phenomenon known as Barren Plateaus (BP). Initially, we did a brief review on VQAs, that are one of the most promising quantum algorithms for the NISQ era. We also did a quick analysis of the training process of VQAs using the gradient descent method. We addressed two problems that the optimization of parameters, using the optimization rule given in Eq. \eqref{eq:gradient}, involves. We related BPs with one of these problems and we showed  that the use of a DNNs is able to avoid the BPs. 
It is worthwhile observing that these positive results of our method are not due to a possible initialization with low values of cost function. In order leave no doubts about this issue, in Appendix B we show plots of the cost function against the number of training epochs. Besides, the success of our method is not due to the classical neural network producing a parametrization that gives a quantum circuit close to the identity matrix, as we show in Appendix C. 

We also showed how the architecture of the DNNs affected the results, and we concluded that for the three models used the behavior was fairly similar. 
However, we must make two observations, the first is related to the choice of the activation function. During the numerical simulations, we saw that only the use of the hyperbolic tangent activation function was able to optimize the parameters to obtain $\pmb{\theta}^ {*}$. The second point is related to the number of hidden layers used in the DNN and their dimensions. As mentioned above, in the performed simulations these numbers did not influence significantly the results. 
However, as in the NISQ era the number of qubits we have access to is limited and the parameterization depth of $U$ is also limited by noise, currently we are not able to determine if for large VQAs this non-dependence will still hold.

Some methods have already been proposed in order to mitigate BPs. Among these methods, we can mention parameter initialization strategies \cite{BR_initialization_strategy}, which consists of initializing the parameters in such a way that the parameterization $U$ acts as an identity matrix. However, with this kind of method, one manages to avoid BPs when initializing the parameterization, but at no time during the optimization of the parameters one can guarantee that the circuit will not suffer from BPs. As we can see with our simulations, this is not the case for our method, that manages to mitigate BPs during training. Another method for alleviating BPs was reported in Ref. \cite{BR_LSTM}. This method is similar to ours, where a classical network is used for proposing the initial parameters. But in that work, the authors used LSTM neural networks. In addition, due to the way these networks interact with the quantum circuit, the number of times that the quantum circuit must be run to obtain all the derivatives necessary for the  parameters optimization is much larger than for our method. Finally, another strategy used for dealing with BPs is the correlation of parameters \cite{BR_Large_gradients_via_correlation}. A consequence of correlating the parameters is that the expressiveness of the parameterization can decrease. In fact, in Ref. \cite{BR_expressibility} a relationship between expressiveness and BPs was obtained, with the greater the expressiveness the greater is the effect of the BPs. In Ref. \cite{BR_Large_gradients_via_correlation}, correlation of parameters was used to decrease the expressiveness of the parameterization. Our method, at first, does not affect the expressiveness. But we must observe that the parameters obtained at the end of the classical neural network are correlated through the hidden layers. So, a more detailed study should be carried out in the future in this regard.

Finally, one could argue that for large VQAs the method proposed in this article would not be viable due to the high number of variables involved. However, we should note that the most modern models of DNNs, in particular models dedicated to natural language processing, are constructed using a large number of variables, reaching billions of parameters. So, this argument is no longer valid. Furthermore, it should be noted that the number of epochs required for large optimizations of VQAs without using DNNs on the initial data would be so high as to make the use of VQAs impractical.

\begin{acknowledgments}
This work was supported  by the Foundation for Research Support of the State of Rio Grande do Sul (FAPERGS), by the National Institute for the Science and Technology of Quantum Information (INCT-IQ), process 465469/2014-0, and by the National Council for Scientific and Technological Development (CNPq), process 309862/2021-3.
\end{acknowledgments}

\vspace{0.3cm}

\textbf{Data availability} The Pennylane/Pytorch code used for implementing the simulations to obtain the data used in this article is available upon request to the authors.

\section{Appendix A}
\label{apendice}
For our demonstrations, we will use the following identities:
\begin{lemma}
\label{lemma_1}
Let $\{W_{y} \}_{y\in Y} \subset U(d)$ form a unitary t-design with $t \geq  1$, and let $A, B : H_{w} \rightarrow H_{w}$ be arbitrary linear
operators. Then
\begin{equation}
    \int d\mu(W)Tr[WAW^{\dagger}B] = \frac{Tr[A]Tr[B]}{d}. \label{eq:identity_1}
\end{equation}
\end{lemma}

\begin{lemma}
\label{lemma_2}
Let $\{W_{y} \}_{y\in Y} \subset U(d)$ form a unitary t-design with $t \geq  2$, and let $A, B : H_{w} \rightarrow H_{w}$ be arbitrary linear
operators. Then
\begin{equation}
    \begin{split}
    \int d\mu(W)Tr[WAW^{\dagger}BWCW^{\dagger}D] = \\ = \frac{ Tr[A]Tr[C]Tr[BD]+Tr[AC]Tr[B]Tr[D]  }{d^{2}-1} \\ - \frac{ Tr[AC]Tr[BD]+Tr[A]Tr[B]Tr[C]Tr[D] }{d(d^{2}-1)}. \label{eq:identity_2}
    \end{split}
\end{equation}
\end{lemma}

\begin{lemma}
\label{lemma_3}
Let $\{W_{y} \}_{y\in Y} \subset U(d)$ form a unitary t-design with $t \geq  2$, and let $A, B, C, D : H_{w} \rightarrow H_{w}$ be arbitrary linear
operators. Then
\begin{equation}
    \begin{split}
    \int d\mu(W)Tr[WAW^{\dagger}B]Tr[WCW^{\dagger}D] = \\ = \frac{ Tr[A]Tr[B]Tr[C]Tr[D]+Tr[AC]Tr[BD] }{d^{2}-1} \\ - \frac{ Tr[AC]Tr[B]Tr[D]+Tr[A]Tr[C]Tr[BD] }{d(d^{2}-1)}, \label{eq:identity_3}
    \end{split}
\end{equation}
\end{lemma}
where $d=2^{n}$. For more details on the proofs of these lemmas, see Refs. \cite{Harr_measure_1,Harr_measure_2}.

\subsection{Derivation of the cost function}
Let us consider
\begin{equation}
    U(\pmb{\theta}) = U_{L}U_{R} \label{eq:demo_1}
\end{equation}
with
\begin{equation}
    U_{R} = \prod_{i=1}^{k}U(\pmb{\theta}_{i})W_{i}  \label{eq:demo_2_R}
\end{equation}
and
\begin{equation}
    U_{L} = \prod_{i=k+1}^{L}U(\pmb{\theta}_{i})W_{i}.
    \label{eq:demo_2_L}
\end{equation}

The derivative of Eq. \eqref{eq:demo_1} with respect to a parameter $\theta_{k}$ will be
\begin{equation}
    \partial_{k}U  = U_{L}\partial_{k}U_{R} = U_{L}\big(\mathbb{I}_{k}\otimes (-i/2)\sigma \big)U_{R}. 
    \label{eq:demo_3}
\end{equation}
In Eq. \eqref{eq:demo_3}, the term $\mathbb{I}_{k}$ indicates that the identity operator will be applied to all qubits except for the qubit where the gate with parameter $\theta_{k}$ acts on.

Considering the cost function defined in Eq. \eqref{eq:cost_function_geral}, we have that the derivative with respect to $\theta_{k}$ is given by
\begin{equation}
    \partial_{k}C = Tr[H ((\partial_{k}U)\rho U^{\dagger} + U\rho (\partial_{k}U^{\dagger})  )].
\end{equation}
Then, using Eqs. \eqref{eq:demo_1} and \eqref{eq:demo_3}, we get
\begin{equation}
    \begin{split}
    \partial_{k}C = Tr\bigg[H \bigg\{ U_{L}\big(\mathbb{I}_{k}\otimes \bigg(\frac{-i}{2}\bigg)\sigma \big)U_{R} \rho \big( U_{L}U_{R} \big)^{\dagger} + \\ U_{L}U_{R}\rho  \bigg(U_{L}\big(\mathbb{I}_{k}\otimes \bigg(\frac{-i}{2}\bigg)\sigma \big)U_{R}\bigg)^{\dagger}   \bigg\}\bigg]. \label{eq:demo_4}
    \end{split}
\end{equation}
Rearranging the terms in Eq. \eqref{eq:demo_4}, we get
\begin{equation}
    \partial_{k}C = \frac{i}{2}Tr\big[ [U_{R}\rho U_{R}^{\dagger}, \mathbb{I}_{k}\otimes \sigma ] U_{L}^{\dagger}HU_{L} \big]. \label{eq:demo_5}
\end{equation}

\subsection{Average value of partial derivative}
To prove average value of partial derivative we will consider three scenarios. The first is when only $U_{R}$ is a 2-design, the second is when only $U_{L}$ is a 2-design, and the third is when both $U_{R}$ and $U_{L}$ are 2-designs.

For the first case, from Eq. \eqref{eq:demo_5} we get
\begin{equation}
    \begin{aligned}
        \langle \partial_{k}C \rangle _{U_{R}} &= \frac{i}{2} \int d\mu(U_{R}) Tr\big[ [U_{R}\rho U_{R}^{\dagger}, \mathbb{I}_{k}\otimes \sigma ] U_{L}^{\dagger}HU_{L} \big]  \\
        & =  \frac{i}{2} \int d\mu(U_{R})  Tr\bigg[ U_{R}\rho U_{R}^{\dagger} (\mathbb{I}_{k}\otimes \sigma) U_{L}^{\dagger}HU_{L}  \bigg] \\
        & - \frac{i}{2} \int d\mu(U_{R})  Tr\bigg[ U_{R}\rho U_{R}^{\dagger}U_{L}^{\dagger}HU_{L} (\mathbb{I}_{k}\otimes \sigma)  \bigg] \\
        & = \frac{i}{2d} Tr[\rho]Tr[(\mathbb{I}_{k}\otimes \sigma) U_{L}^{\dagger}HU_{L}]\\
        & - \frac{i}{2d} Tr[\rho]Tr[U_{L}^{\dagger}HU_{L}(\mathbb{I}_{k}\otimes \sigma)] \\
        & = 0.
    \end{aligned}
\end{equation}
In the second term of the second equality, and in the last equality, the cyclic property of the trace operator was used. To obtain the third equality, we use the Lemma \ref{lemma_1}. 

For the second case, we just use the cyclic property of the trace operator in Eq.\eqref{eq:demo_5} and integrate over $U_{L}$. So we get
\begin{equation}
    \langle \partial_{k}C \rangle _{U_{L}} = 0.
\end{equation}

The third and last case, when $U_{R}$ and $U_{L}$ are 2-designs, is an extension of the first and second cases. So
\begin{equation}
    \langle \partial_{k}C \rangle _{U} = \langle \partial_{k}C \rangle _{U_{R}U_{L}} = 0.
\end{equation}
This proves that the average value of partial derivative is $\langle \partial_{k}C \rangle _{U} = 0.$

\clearpage
\section{Appendix B}
\begin{figure}[h]
    \centering
    \includegraphics[width=9cm]{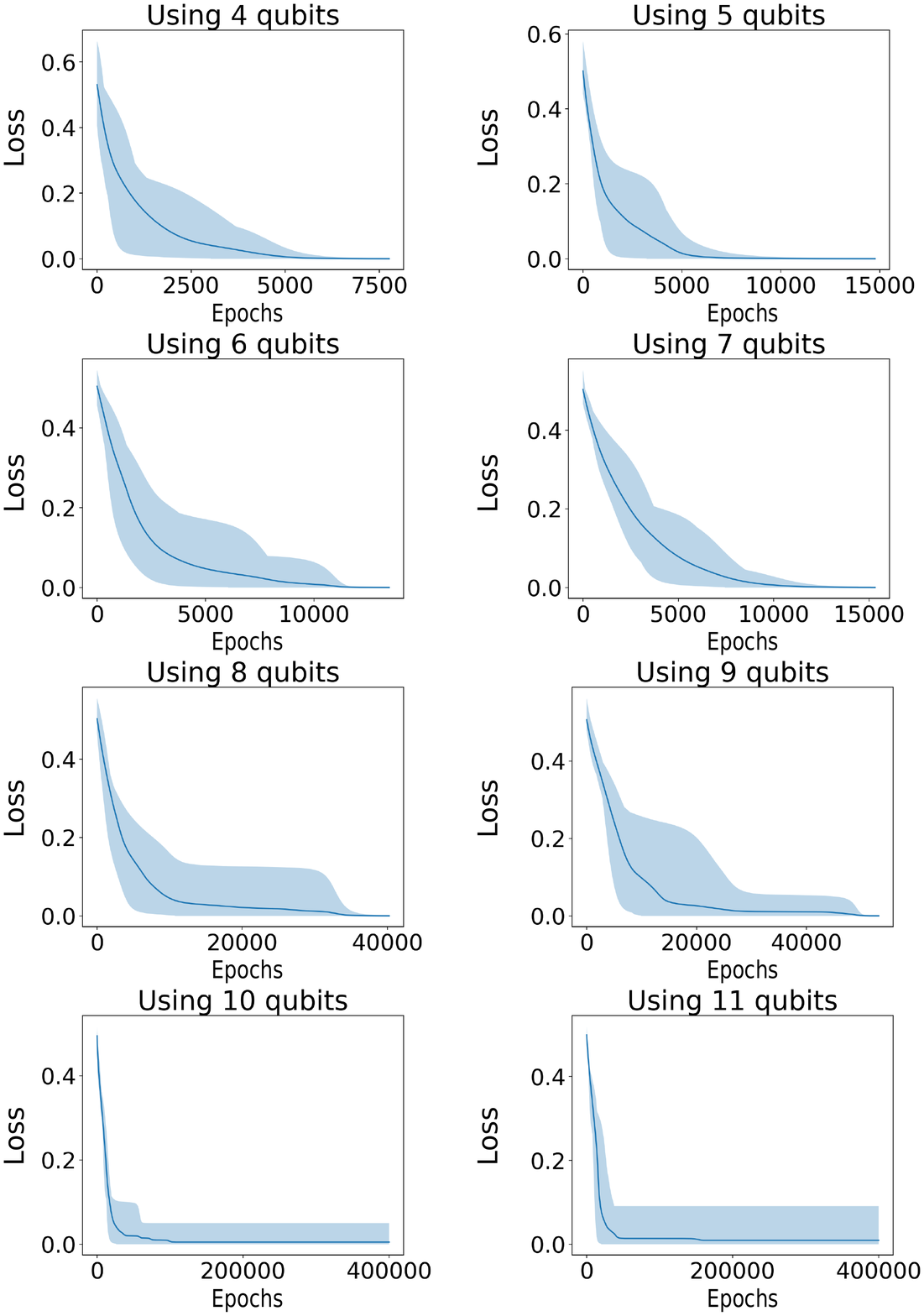}
    \caption{Graphs of the cost function against the number of epochs for the model Net. }
    \label{fig:AB_net}
\end{figure}

\begin{figure}[h]
    \centering
    \includegraphics[width=9cm]{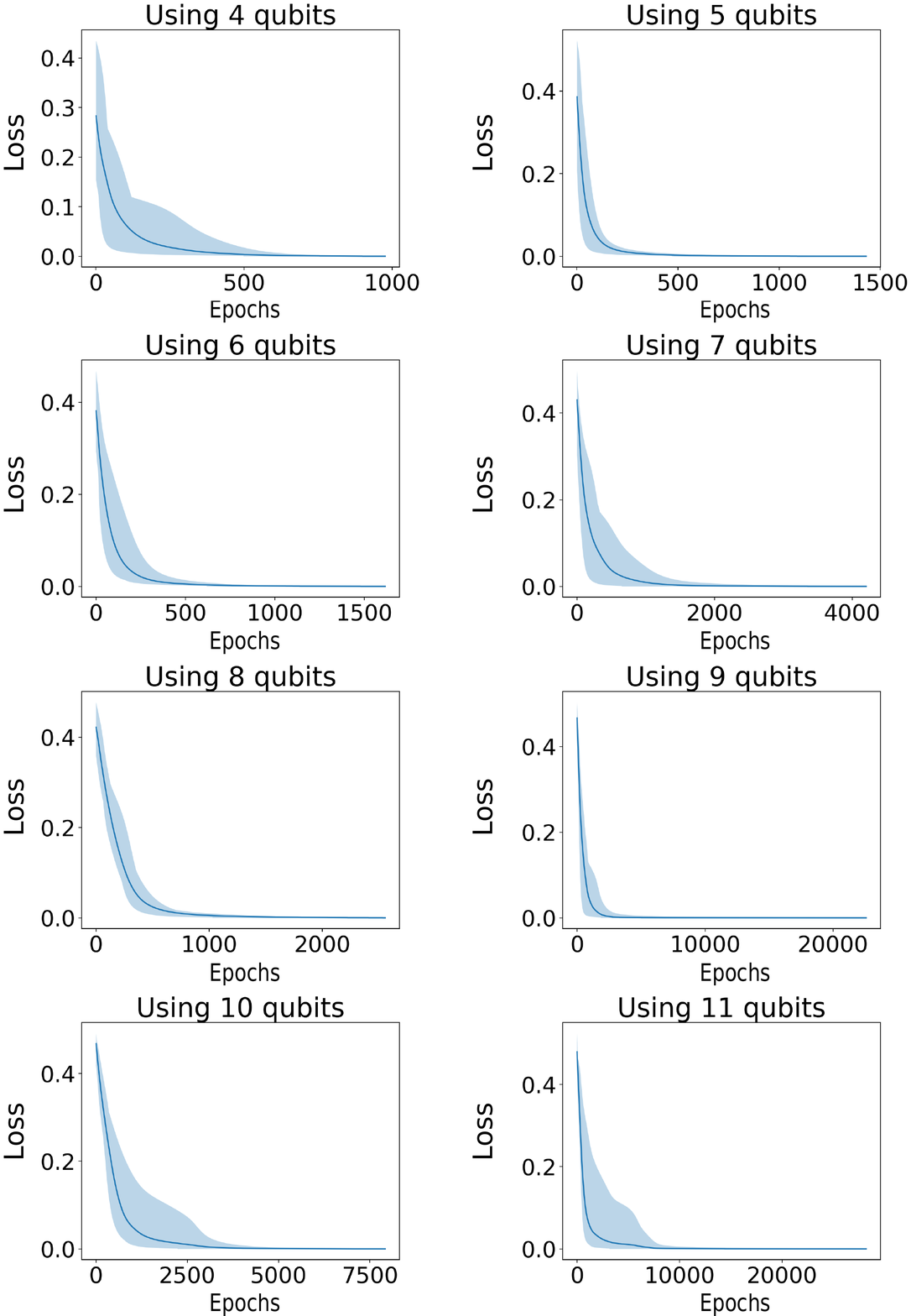}
    \caption{Cost function against the number of epochs for the Model 1.}
    \label{fig:AB_model1}
\end{figure}

\begin{figure}[h]
    \centering
    \includegraphics[width=9cm]{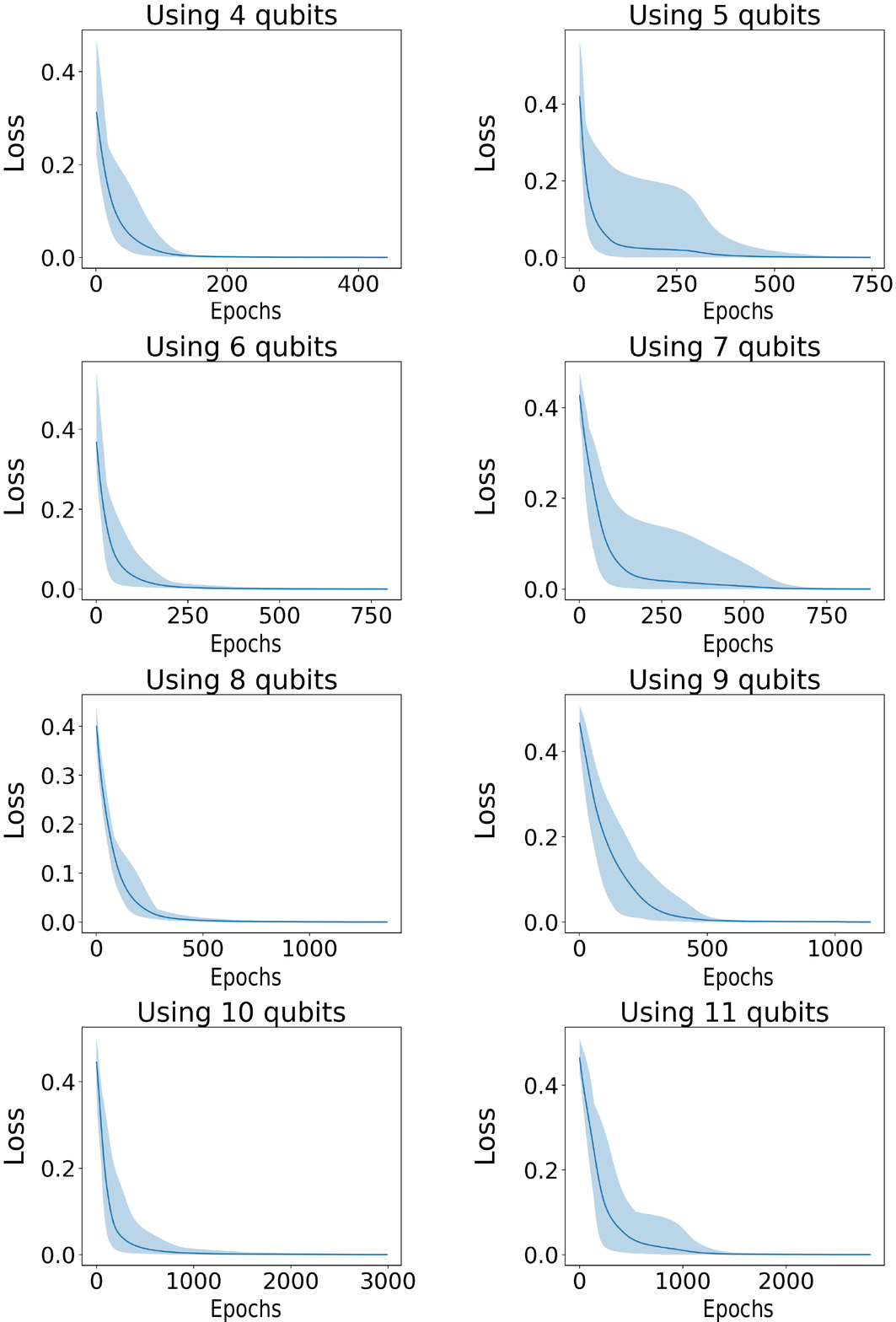}
    \caption{Cost function against the number of epochs for the Model 2.}
    \label{fig:AB_model2}
\end{figure}

\begin{figure}[h]
    \centering
    \includegraphics[width=9cm]{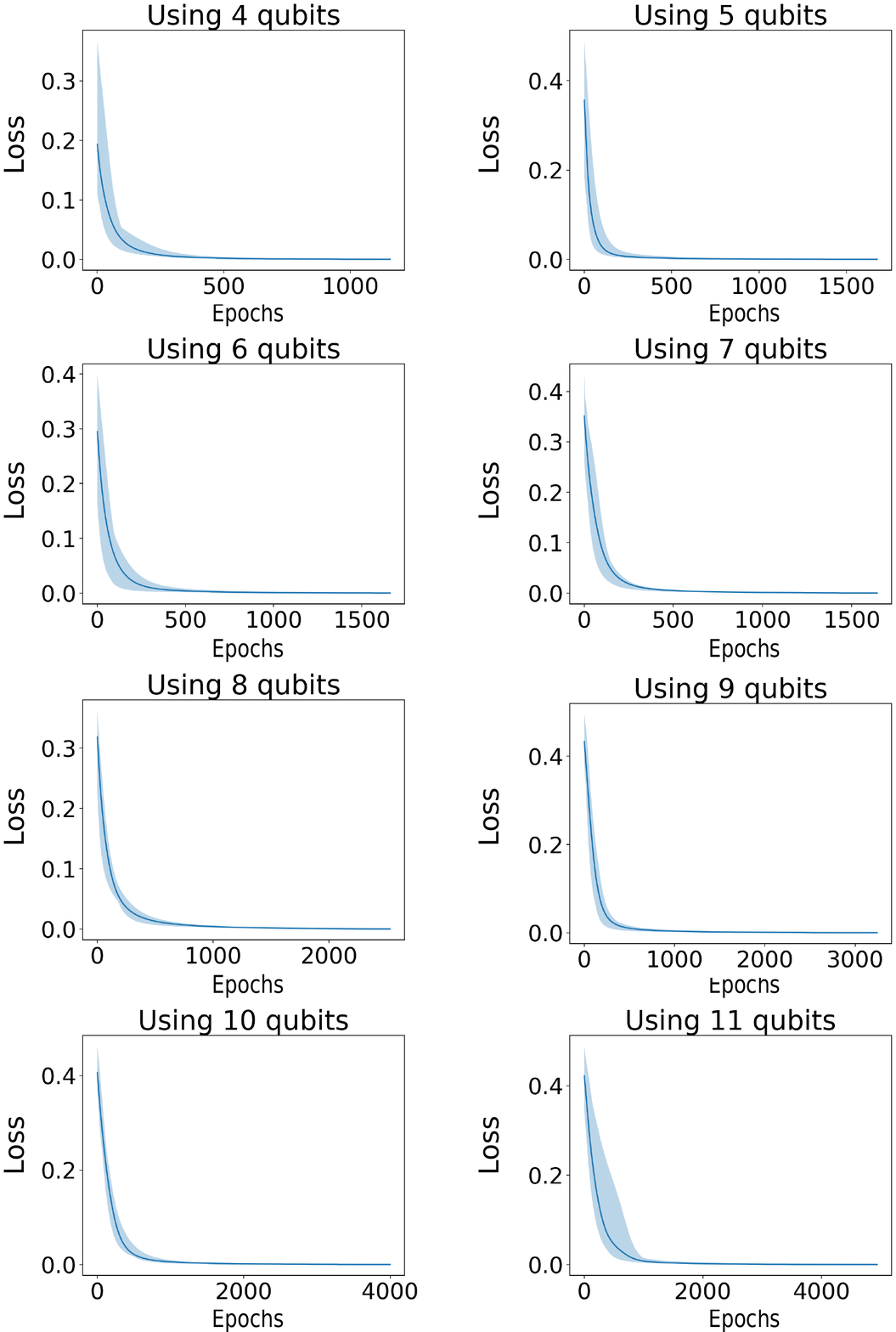}
    \caption{Cost function against the number of epochs for the Model 3.}
    \label{fig:AB_model3}
\end{figure}

In order to show that the positive results of our method are not due to initialization with low values of cost function, here we show plots of the cost function against the number of training epochs.
The graphs show the average, maximum, and minimum values behavior of the cost function. In Figs. 
\ref{fig:AB_net}, \ref{fig:AB_model1}, \ref{fig:AB_model2}, and \ref{fig:AB_model3} 
are shown the results for the Model Net, Model 1, Model 2, and Model 3, respectivelly. In all cases, the parameterization depth grows linearly with the number of qubits.


\clearpage

\section{Appendix C}

In this appendix, we show that the success of our method is not due to the classical neural network producing a parametrization that gives a quantum circuit close to the identity matrix. We present here the behavior of the norm of the difference between the initial state, $|\phi_{i}\rangle$, which is obtained using definition \ref{df:2} with $\pmb{x}_{i} = (\frac{\pi}{4}, \frac{\pi}{4}, \dots,\frac{\pi}{4})$, and the state after applying the parameterization using the $\pmb{\theta}$ data from the classical neural network, $|\phi_{f}\rangle=U(\pmb{\theta})|\phi_{i}\rangle$:
\begin{equation}
   \mu = || |\phi_{f}\rangle - |\phi_{i}\rangle  ||=|| (U(\pmb{\theta}) - \mathbb{I} )|\phi_{i}\rangle  ||.
\end{equation}
We this we can analyze if the parameters $\pmb{\theta}$ given by the classical neural network are such that the quantum circuit corresponds to the identity matrix, because in this case, as seen in Ref. \cite{BR_initialization_strategy}, there would be no BPs.

\begin{figure}[h]
    \centering
    \includegraphics[width=8cm]{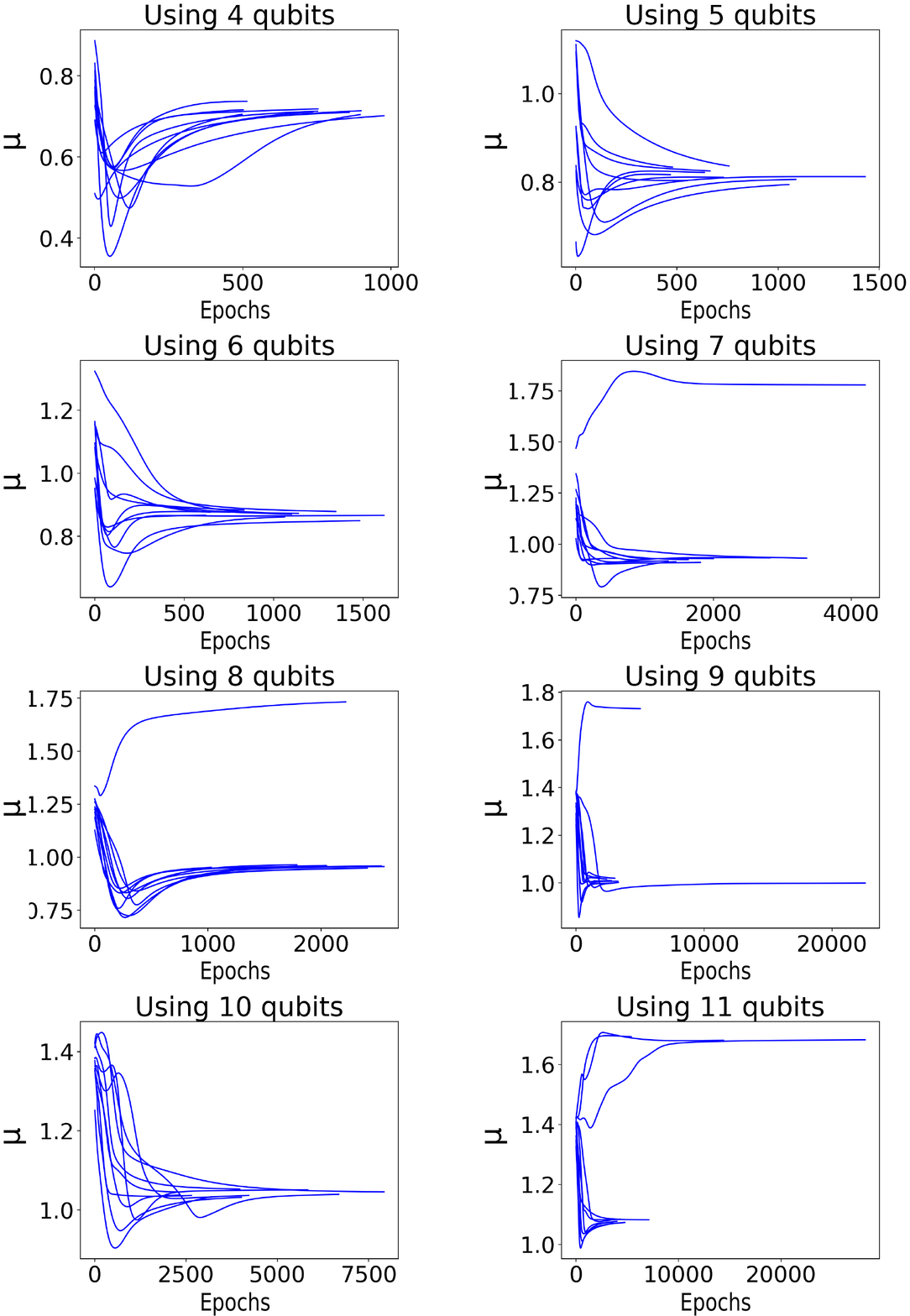}
    \caption{Graphs of the norm    of the difference between the state after the parametrization, $U(\vec{\theta})|\phi_{i}\rangle$, and the initial state, $|\phi_{i}\rangle$, for the Model 1.}
    \label{fig:norma_Model1}
\end{figure}

\begin{figure}[h]
    \centering
    \includegraphics[width=8cm]{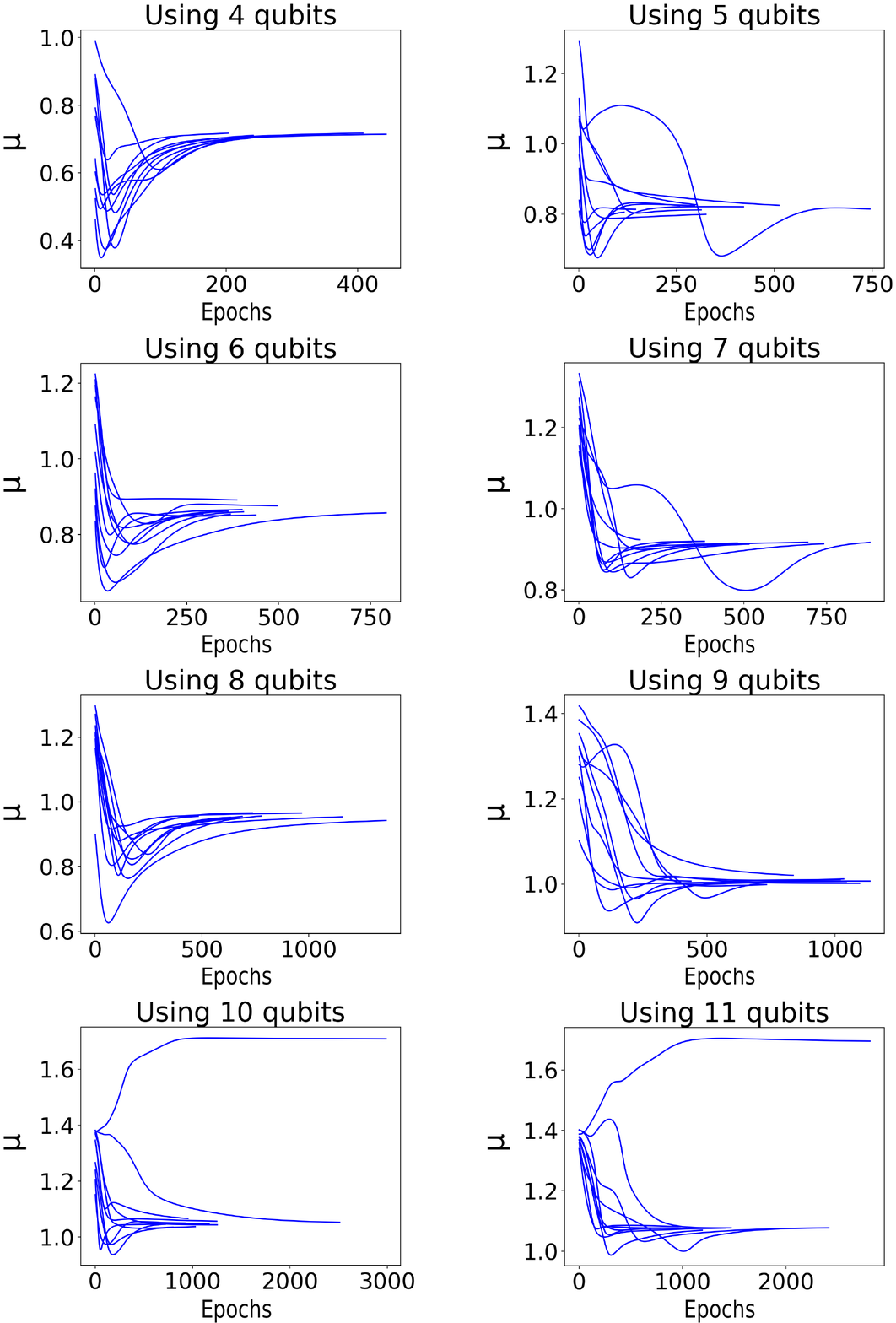}
    \caption{Graphs of the norm 
    of the difference between the state after the parametrization, $U(\vec{\theta})|\phi_{i}\rangle$, and the initial state, $|\phi_{i}\rangle$, for the Model 2.}
    \label{fig:norma_Model2}
\end{figure}

\begin{figure}[h]
    \centering
    \includegraphics[width=8cm]{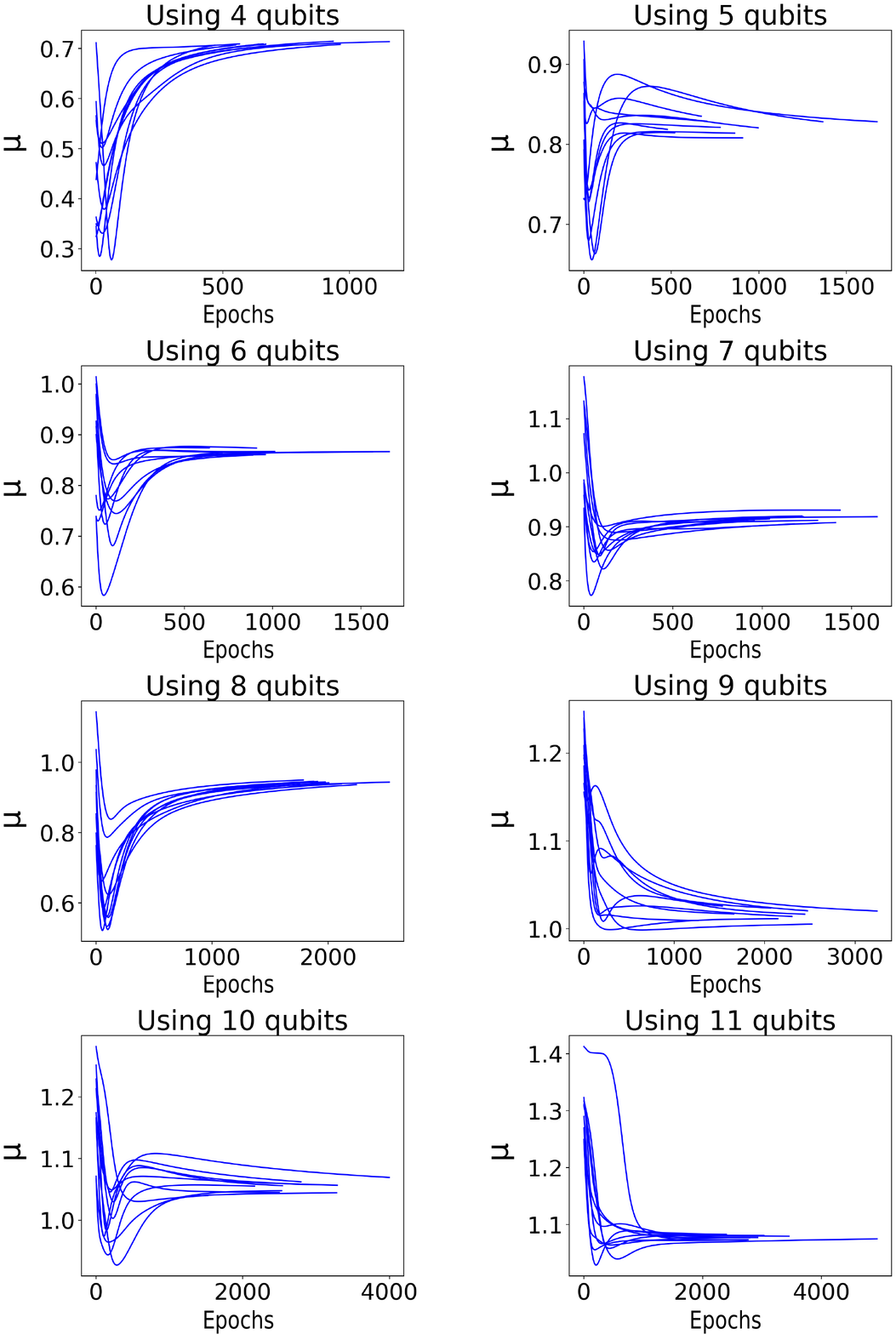}
    \caption{Graphs of the norm 
    of the difference between the state after the parametrization, $U(\vec{\theta})|\phi_{i}\rangle$, and the initial state, $|\phi_{i}\rangle$, for the Model 3.}
    \label{fig:norma_Model3}
\end{figure}

We can see in Figs. \ref{fig:norma_Model1}, \ref{fig:norma_Model2} and \ref{fig:norma_Model3}  that the parameters obtained by the classical network when applied to the quantum circuit do not act as an identity matrix. Here also the parameterization depth grows linearly with the number of qubits. So the classical neural network does not alleviate the BP because the parameterization obtained has a behavior similar to an identity matrix. With this we can see that there is another phenomenon behind this behavior. Therefore, our results are from a different origin when compared with the ones obtained in Ref. \cite{BR_initialization_strategy}.

\clearpage

\begin{thebibliography}{4}

\bibitem{quantum_simulation} S. Lloyd, Universal quantum simulators, Science 273, 1073 (1996).

\bibitem{linear_system} A. W. Harrow, A. Hassidim, and S. Lloyd, Quantum algorithm for linear systems of equations, Phys. Rev. Lett. 103, 150502 (2009).

\bibitem{quantum_nlp_1} B. Coecke, G. de Felice, K. Meichanetzidis, and A. Toumi, Foundations for near-term quantum natural language processing, 
\url{https://doi.org/10.48550/arXiv.2012.03755} (2020).

\bibitem{quantum_nlp_2}  K. Meichanetzidis, S. Gogioso, G. de Felice, N. Chiappori, A. Toumi, and B. Coecke, Quantum natural language processing on near-term quantum computers, EPTCS 340, 213 (2021).

\bibitem{drug_discovery} Y. Cao, J. Romero, and A. Aspuru-Guzik, Potential of quantum computing for drug discovery, IBM Journal of Research and Development 62, 6 (2018).

\bibitem{VQA} M. Cerezo et al., Variational quantum algorithms, Nature Reviews Physics 3, 625 (2021).

\bibitem{BR_cost_Dependent} M. Cerezo, A. Sone, T. Volkoff, L. Cincio, and P. J. Coles, Cost function dependent barren plateaus in shallow parametrized quantum circuits, Nature Communications 12, 1791 (2021).

\bibitem{BR_Entanglement_devised_barren_plateau_mitigation} T. L. Patti, K. Najafi, X. Gao, and S. F. Yelin, Entanglement devised barren plateau mitigation, Phys. Rev. Research 3, 033090 (2021).

\bibitem{BR_Entanglement_induced_barren_plateaus} C. O. Marrero, M. Kieferov\'a, and N. Wiebe, Entanglement-induced barren plateaus, PRX Quantum 2, 040316 (2021).

\bibitem{BR_expressibility} Z. Holmes, K. Sharma, M. Cerezo, and P. J. Coles, Connecting ansatz expressibility to gradient magnitudes and barren plateaus, PRX Quantum 3, 010313 (2022).

\bibitem{BR_noise} S. Wang et al., Noise-induced barren plateaus in variational quantum algorithms, Nature Communications 12, 6961  (2021).

\bibitem{BR_gradientFree} A. Arrasmith, M. Cerezo, P. Czarnik, L. Cincio, and P. J. Coles, Effect of barren plateaus on gradient-free optimization,  Quantum 5, 558 (2021).

\bibitem{BR_initialization_strategy} E. Grant, L. Wossnig, M. Ostaszewski, and M. Benedetti, An initialization strategy for addressing barren plateaus in parametrized quantum circuits, Quantum 3, 214 (2019).

\bibitem{BR_Large_gradients_via_correlation} T. Volkoff and P. J. Coles, Large gradients via correlation in random parameterized quantum circuits, Quantum Science and Technology 6, 025008 (2021).

\bibitem{BR_LSTM} G. Verdon et al, Learning to learn with quantum neural networks via classical neural networks, \url{
https://doi.org/10.48550/arXiv.1907.05415} (2019).

\bibitem{BR_layer_by_layer} A. Skolik, J. R. McClean, M. Mohseni, P. van der Smagt, and M. Leib, Layerwise learning for quantum neural networks, Quantum Machine Intelligence 3, 5 (2021).


\bibitem{data_encoder_1} R. LaRose and B. Coyle, Robust data encodings for quantum classifiers, Phys. Rev. A 102, 032420 (2020).

\bibitem{data_encoder_2} A. P\'erez-Salinas, A. Cervera-Lierta, E. Gil-Fuster, and J. I. Latorre, Data re-uploading for a universal quantum classifier, Quantum 4, 226 (2020).

\bibitem{data_encoder_3} M. Schuld, R. Sweke, and J. J. Meyer, Effect of data encoding on the expressive power of variational quantum-machine-learning models, Phys. Rev. A 103, 032430 (2021).

\bibitem{expressibility_1} S. Sim, P. D. Johnson, and A. Aspuru-Guzik, Expressibility and entangling capability of parameterized quantum circuits for hybrid quantum classical algorithms, Advanced Quantum Technologies 2, 1900070 (2019).

\bibitem{expressibility_2} K. Nakaji and N. Yamamoto, Expressibility of the alternating layered ansatz for quantum computation, Quantum 5, 434 (2021).

\bibitem{Expressibility_barren_plateaus} Z. Holmes, K. Sharma, M. Cerezo, and P. J. Coles, Connecting ansatz expressibility to gradient magnitudes and barren plateaus, PRX Quantum 3, 010313 (2022).

\bibitem{Barren_Plateaus_1} J. R. McClean, S. Boixo, V. N. Smelyanskiy, R. Babbush, and H. Neven, Barren plateaus in quantum neural network training landscapes, Nature Communications 9, 4812 (2018).

\bibitem{pytorch_ref} A. Paszke et al., Pytorch: An imperative style, high-performance deep learning library, \url{
https://doi.org/10.48550/arXiv.1912.01703} (2019).

\bibitem{Harr_measure_1} B. Collins and P. \'Sniady, Integration with respect to the Haar measure on unitary, orthogonal and symplectic group, Commun. Math. Phys. 264, 773 (2006).

\bibitem{Harr_measure_2} Z. Pucha\l a. and J. A. Miszczak, Symbolic integration with respect to the Haar measure on the unitary group, Bull. Pol. Acad. Sci.-Tech. Sci. 65, 1 (2017).


\end{thebibliography}
\end{document}